\documentclass[reprint, prmaterials, amsfonts, amssymb, amsmath, aps, floatfix, article]{revtex4-2}
\usepackage[pdfusetitle]{hyperref}
\usepackage{graphicx}
\usepackage{dcolumn}
\newcolumntype{d}[1]{D{.}{.}{#1}}
\usepackage{bm}
\usepackage{longtable}
\usepackage{mathtools}
\usepackage{ulem}

\usepackage{tcolorbox}
\tcbuselibrary{breakable, skins}

\NewTColorBox{scratch}{ }{
    enhanced,
    breakable,
    size=minimal,
    colback=transparent!0,
    colframe=transparent!0,
    overlay={
        \draw (frame.south west)
            -- (frame.north east);
        \draw (frame.north west)
            -- (frame.south east);
    }
}

\usepackage{xspace}
\usepackage{xpunctuate}
\DeclareRobustCommand\etal{\xperiodafter{\emph{et al}}}

\hypersetup{
  colorlinks=true,
  linkcolor=blue,
  filecolor=blue,
  citecolor=blue,
  urlcolor=blue,
}

\usepackage{xcolor}

\usepackage{float}

\NewDocumentEnvironment{FullWidth}{ +b }{
    \twocolumn[{#1}]%
}{}

\begin{document}

\title{Deep-lying semi-Dirac fermions in hexagonal close-packed cadmium }

\author{Alaska Subedi} 
\affiliation{CPHT, CNRS, \'Ecole polytechnique, Institut Polytechnique
  de Paris, 91120 Palaiseau, France} 

\author{Kamran Behnia}
\affiliation{Laboratoire de Physique et d'\'Etude des Mat\'eriaux
  (ESPCI - CNRS - Sorbonne Universit\'e), Universit\'e Paris Sciences
  et Lettres, 75005 Paris, France}

\date{\today}

\begin{abstract}

Semi-Dirac fermions are massless in one direction and massive in the
perpendicular directions. Such quasiparticles have been proposed in
various contexts in condensed matter.  Using first principles
calculations, we identify a pair of semi-Dirac bands anti-crossing at $-3$ eV below the Fermi level in the electronic structure of hexagonal close-packed (hcp) cadmium. The linear out-of-plane dispersion is kept up to the Fermi level. We demonstrate that the dichotomy between the linear and quadratic dispersions is driven by an orientation-sensitive hybridization between the $s$ and $p_z$ orbitals. The upper semi-Dirac band produces a lens-shaped nonellipsoidal Fermi sheet whose cross-section area has a $k$-dependence that is in excellent agreement with the experimentally measured period of Sondheimer oscillations.

\end{abstract}

\maketitle

\section{Introduction}

Linearly dispersive electronic bands can give rise to massless analogues of relativistic particles in condensed matter systems. Such Dirac fermions lead to intriguing experimental consequences like the $E_N \propto \sqrt{N B}$ energy spectrum of the Landau levels, half-integer anomalous quantum Hall effect, and Klein tunneling \cite{Katsnelson2020}.  Observation of these phenomena in monolayer graphene \cite{Geim2007}, which has Dirac cones at its Brillouin zone corners, has motivated the search for other such systems.

Semi-Dirac fermions exhibiting direction-dependent massless and massive behaviors have attracted much attention during the last two decades.
They were proposed to exist in the $A$ phase of $^3$He \cite{Volovik2001}, in a graphenelike model subjected to a magnetic field \cite{Dietl2008}, VO$_2$ layers confined within TiO$_2$ \cite{Pardo2009}, strained silicene oxide \cite{Zhong2017}, and tetragonal perovskite oxides with the $I4/mcm$ space group \cite{Mohanta2021}. 
The merging-of-Dirac-points scenario provides a unifying topological picture for this phenomenon \cite{Montambaux2009a,Montambaux2009b}, with Bloch–Zener dynamics across the transition analyzed in \cite{Lim2012}. A magnetic-flux analogue appears in the Hofstadter spectrum \cite{Delplace2010}, and transport and criticality near the merging transition were analyzed in Refs.~\cite{Adroguer2016,Isobe2016}.
While their observations have been reported in microwave a analogue of a graphenelike lattice \cite{Bellec2013}, honeycomb lattice of micropillars \cite{Real2020}, and cold atoms \cite{Tarruell2012}, they have been more elusive in solids. Recently, a semi-Dirac dispersion leading to a $[(N+\frac{1}{2})B]^{2/3}$ scaling of the Landau levels \cite{Dietl2008,Banerjee2009} has been reported in ZrSiS \cite{Shao2023}.


Generally, linear dispersion is expected when two nonbonding bands cross at a Dirac point situated away from the Brillouin zone center. Semi-Dirac bands in (TiO$_2$)$_5$/(VO$_2$)$_3$ arise in this way \cite{Pardo2009}. Another avenue is the inversion of the relative positions of bands associated with $s$ and $p$ orbitals. In HgTe-CdTe quantum wells, such an inversion occurs at a critical thickness leading to a Dirac point at the Brillouin zone center \cite{Bernevig2006}. Given that $s$-$p$ hybridization is uniform along all three axes in cubic zinc blende HgTe, however, a semi-Dirac analogue of this mechanism has  been missing.

A possible avenue to obtain semi-Dirac bands is to  explore cases with direction-dependent $s$-$p$ hybridization.  Metals in column twelve of the periodic table are promising, as their mobile electrons originate from $s$ and $p$  orbitals rather than $d$ orbitals.  Furthermore, unlike column eleven elements (the noble metals) with a nominal electronic configuration $s^1$, column twelve metals are nominally $s^2$. The additional electron in column 12 metals favors the hcp structure, where the  $p_z$ orbital hybridizes differently with the $s$ orbital compared to the $p_x$ and $p_y$ orbitals.

In this paper, we use first principles calculations to show that hcp
Cd hosts a pair of semi-Dirac bands anti-crossing at $-3$ eV below the Fermi level.
The bands' dispersion is parabolic in plane and linear out of the
plane over an extended distance in the reciprocal space that includes
the Brillouin zone center and the Fermi level crossing.  The two bands
have mixed $s$ and $p_z$ character in the out-of-plane direction and
unmixed $s$ or $p_z$ character in the in-plane directions.  Thus, the
direction-dependent mixing of the $s$ and $p_z$ characters is
responsible for the presence of the semi-Dirac bands.  Comparison with
hypothetical hcp Ag shows that the relative positions of the $s$- and
$p_z$-like bands control the direction-dependent mixing.  We show that
in such a hypothetical hcp metal with a single $s$ electron, there is
a large gap between $s$-like and $p_z$-like states.  Consequently,
$s$-$p_z$ hybridization is negligible and the bands are parabolic in
all directions.  We find that tensile strain moves the $s$-like states
higher relative to the $p_z$ states, thereby increasing the $s$-$p_z$
hybridization in the out-of-plane direction.  Semi-Dirac dispersion
occurs near the point of $s$-$p_z$ inversion, in a manner analogous to
that of HgTe-CdTe quantum wells where the band inversion is controlled
by tuning the strength of the effective spin-orbit interaction.  The
Fermi level crossing of the upper semi-Dirac band generates a
non-ellipsoid Fermi pocket, called lens-shaped in the scientific
literature. We find that the calculated derivative of the
cross-section area of this sheet with respect to the out-of-plane wave
vector agrees with the experimental value extracted from the observed
Sondheimer oscillations, corroborating the presence of the calculated
semi-Dirac bands in this material.

\begin{figure}
  \includegraphics[width=\columnwidth]{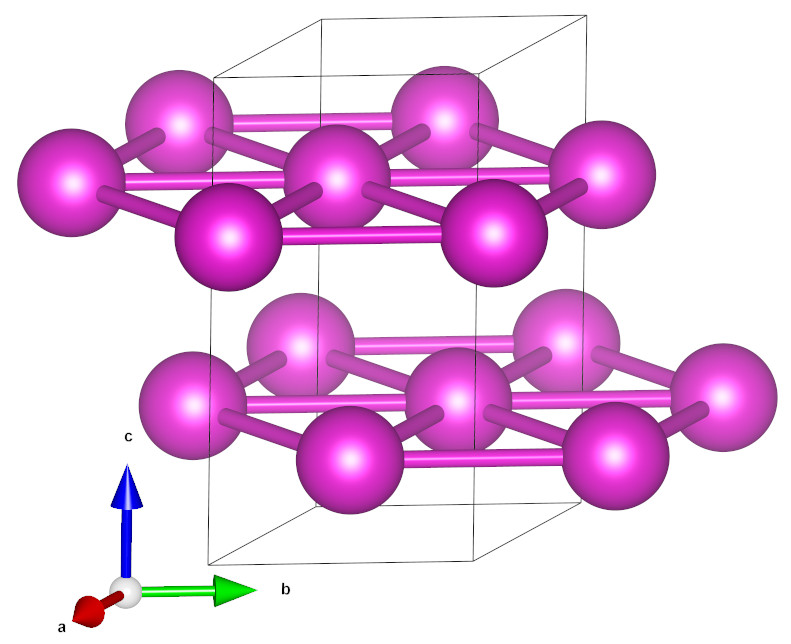}
  \caption{Hexagonal closed packed structure of cadmium. The inter-
  and intra-layer nearest-neighbor Cd-Cd distances are 3.293 and 
  2.979 \AA, respectively. }
  \label{fig:struct}
\end{figure}

\section{Computational details} 

The first principles calculations presented here were performed using
the generalized full-potential method as implemented in the {\sc
  wien2k} package \cite{wien2k}.  We used the local density
approximation (LDA) for the exchange-correlation functional and
muffin-tin radius $R$ of 2.5 a.u. The plane-wave cutoff
$K_{\textrm{max}}$ was set using the condition $RK_{\textrm{max}} =
7$. A $72 \times 72 \times 36$ $k$-point grid was used for the
Brillouin zone integration.  The {\sc skeaf} code was modified inhouse
to output the cross-section area of the Fermi sheet as a function of
the $k_z$ wave vector \cite{skeaf}. The effective mass was obtained by 
fitting the semi-Dirac band along the inplane direction by the polynomial 
$E(k) = E_0 + \hbar^2m k^2/2$ around $\pm0.05$ \AA$^{-1}$ near the Fermi 
level crossing. Results presented in the paper do not include the 
spin-orbit interaction.  However, we verified that inclusion of spin-orbit
interaction does not change the dispersion of the semi-Dirac bands.

The hexagonal close packed structure of cadmium is shown in Fig.~\ref{fig:struct}. 
We used the experimental lattice parameters $a = 2.979$ and $c = 5.617$
\AA\ in all the calculations \cite{Lynch1965}.  The two atoms in the unit cell
are situated at $(1/2,2/3,1/4)$ and $(2/3,1/3,3/4)$.  Note that
the $c/a$ ratio of 1.886 of hcp Cd is much larger than the ideal value
of $\sqrt{8/3} \sim 1.633$ required for the densest packing. The
deviation likely results from the additional complexity of interatomic
bonding due the presence of the $4d$ states near the Fermi level. Contrast 
this to hcp Mg (that does not have $3d$ states), where $c/a = 1.623$.

\begin{figure*}[!ht]
  \includegraphics[width=\textwidth]{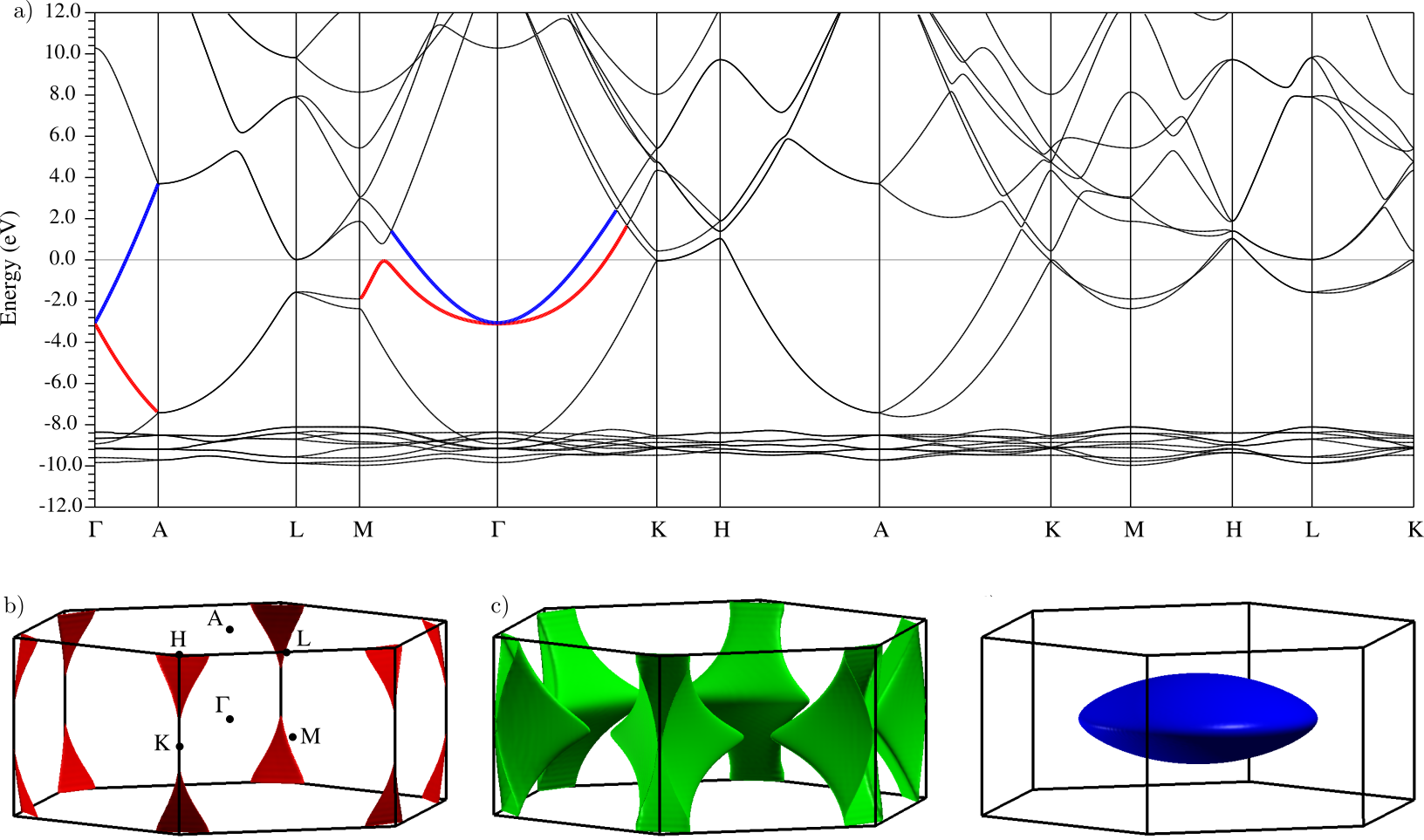}
  \caption{a) Calculated LDA band structure of hcp Cd. The band
    structure features a pair of bands that disperse linearly along
    the out-of-plane direction $\Gamma A$ and quadratically along the
    inplane directions $\Gamma M$ and $\Gamma K$. These bands appear
    to converge near $-3$ eV at $\Gamma$, but they are in fact
    separated by 50 meV. The upper and lower semi-Dirac bands are
    shown in blue and red, respectively.  Calculated Fermi surface of
    hcp Cd consisting of b) hole-type ``cap", c) hole-type ``monster",
    and d) electron-type ``lens" sheets.  The lens sheet derives from
    the upper semi-Dirac band.}
  \label{fig:bandst}
\end{figure*}

\section{Results and Discussion}

The calculated LDA band structure of hcp Cd is shown in Fig.~\ref{fig:bandst}.  
There are ten narrow bands with $4d$ character between $-10$ and $-8$ eV relative 
to the Fermi level.  Dispersive bands with $5s$ and $5p$ characters lie above
and cross the Fermi level.  The calculated Fermi surface of hcp Cd is shown in 
Fig.~\ref{fig:bandst}b-d. It consists of two hole-type sheets known as ``caps'' 
and ``monster'' centered around the six edges of the Brillouin zone and the 
electron-type lens sheet centered at $\Gamma$.  The volume enclosed by these sheets 
are $0.01$ and $0.19$ holes and $0.20$ electrons per cell, respectively.  The 
corresponding carrier concentrations are 0.17, 3.31 and $-3.48 \times 10^{22}$ cm$^{-3}$.

Our results generally agree with those obtained by Stark and Falicov 
using empirical pseudopotential method \cite{Stark1967} and Daniuk \etal 
through self-consistent first principles method \cite{Daniuk1989}, although 
there are differences. All three results show one band crossing the Fermi level 
along $\Gamma A$ and $\Gamma M$, while two bands cross along $\Gamma K$, $H A$, 
and $H L$.  One band in our and Stark and Falicov's calculations
traverses the $K$ point at a saddle point, whereas such a crossing is
missing in that of Daniuk \etal. Furthermore, the doubly degenerate
band that almost touches the Fermi level at the $L$ point in our
calculations lie noticeably higher in the other two results.

A prominent feature of the electronic structure of hcp Cd is the pair of bands 
that linearly disperse away from $-3$ eV in opposite directions along the 
out-of-plane path $\Gamma A$.  In Stark and Falicov's study, these bands 
appear to taper off quadratically when they approach each other at $\Gamma$.  
However, the linear dispersion, although not explicitly noted, is visible in 
the results of Daniuk \textit{et al.}\cite{Daniuk1989}.  Interestingly, these 
bands disperse quadratically along $\Gamma K$ and $\Gamma M$. This implies that 
the electronic states are massless in the out-of-plane direction and massive 
within the plane.  We obtain 0.8$m_e$ for the inplane effective mass by a 
quadratic fit near the Fermi level crossing.  It is also striking that the 
hole-type linearly dispersing band along out-of-plane direction becomes 
electron-like when it shows quadratic dispersion along the in-plane directions.
These pair of bands should accurately be called $1/3$-Dirac 
bands.  However, we follow the convention in the literature and
call them semi-Dirac bands.

\begin{figure}[!htbp]
  \includegraphics[width=\columnwidth]{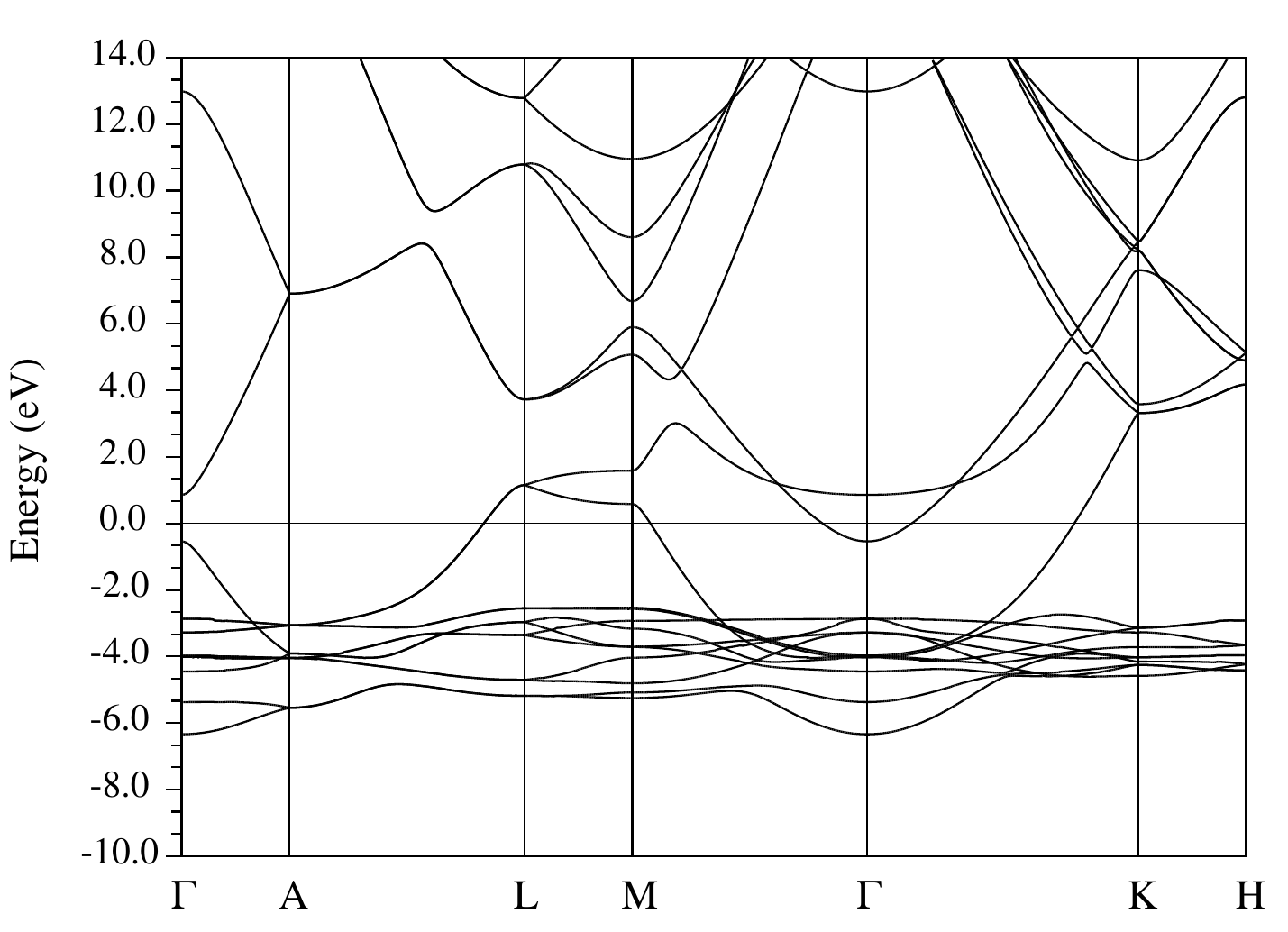}
  \caption{Calculated LDA band structure of a \textit{hypothetical} hcp Ag using the lattice parameters of hcp Cd. The pair of bands that linearly disperse along $\Gamma A$ in hcp Cd disperse quadratically in hypothetical
  hcp Ag. The gap at $\Gamma$ between these bands also increases to
  1.4 eV. }
  \label{fig:bandstAg}
\end{figure}

\begin{figure*}[!htb]
  \includegraphics[width=0.9\textwidth]{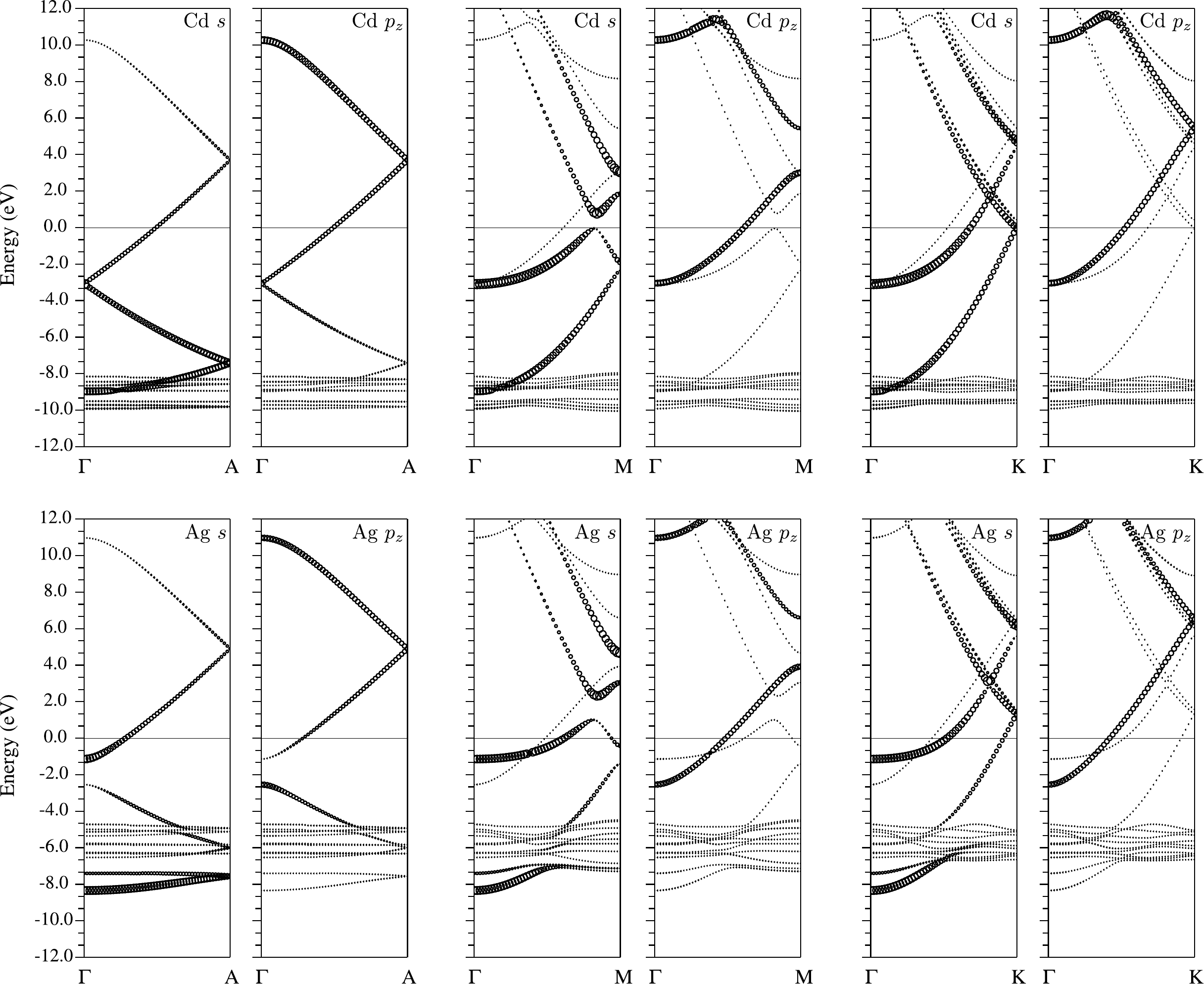}\\
  \caption{Band structures of hcp Cd and hypothetical hcp Ag
    isostructural to hcp Cd as a function of respective $s$ and $p_z$
    orbital characters along out-of-plane $\Gamma A$ and inplane
    $\Gamma M$ and $\Gamma K$ directions.  The semi-Dirac bands in hcp
    Cd show a mixed $s$ and $p_z$ character in the out-of-plane
    direction.  In hcp Ag, the bands disperse quadratically in all
    directions, and the upper band is predominantly $s$-like, while
    the lower band is mainly $p_z$-like.}
  \label{fig:bandch}
\end{figure*}

Semi-Dirac points previously discussed in the literature are
associated with degeneracies due to band crossings \cite{Dietl2008,
  Pardo2009, Banerjee2009, Delplace2010}. In the present case, 
there is a gap of 50 meV  at $\Gamma$ near $-3$ eV where two semi-Dirac 
bands converge even without spin-orbit interaction.  To disentangle 
the role of structure and chemistry behind the origin of these 
semi-Dirac bands, let us perform a gedanken experiment by supposing
that the next-door element in the 
periodic table Ag, instead of being face-centered cubic (its stable 
structure), could be synthesized in the hcp structure
of Cd. 
Fig.~\ref{fig:bandstAg} shows the band structure of such a hypothetical 
hcp Ag with the same structural parameters as that of hcp Cd. We can 
see that the band structures of isostructural hcp Ag and Cd differ by 
more than a rigid shift of the Fermi level even though they lie next 
to each other in the periodic table and their electron count
differs only by one. Hcp Ag no longer exhibits the linearly dispersing
bands along the out-of-plane direction $\Gamma A$.  The gap between
these bands increases to 1.4 eV, and they taper off parabolically as
they approach $\Gamma$.  Therefore, a crossing of nonbonding bands do
not seem to play a role in the appearance of the semi-Dirac bands in
hcp Cd.  Furthermore, the contrasting electronic structures of
isostructural hcp Cd and Ag shows that the semi-Dirac bands do not
arise merely due to the layered hexagonal structure and points out the
essential role played by the chemistry of divalent $5s^2$ electronic
configuration in Cd.



We have plotted the band structures of hcp Cd and Ag showing the
contribution of $s$ and $p_z$ orbital characters along the
out-of-plane and two in-plane directions in Fig.~\ref{fig:bandch} to
understand the changes in band hybridization due to the presence of
an additional $5s$ electron in Cd. Along the out-of-plane $\Gamma A$ 
direction, we can see that the pair of semi-Dirac bands in hcp Cd show 
nearly equal amount of $s$ and $p_z$ characters, especially near 
$\Gamma$. In hcp Ag, in contrast, the upper-lying band shows dominant 
$s$ character, while the $p_z$ character is dominant in the lower-lying 
band.  This is particularly the case close to $\Gamma$, where the 
upper-lying and lower-lying bands are almost solely due to the $s$ 
and $p_z$ states, respectively. Thus, $s$-$p_z$ hybridization along 
one orientation gives rise to a linear dispersion, and the lack of it 
along the two others leads to a quadratic one. This observation is 
further reinforced looking at the orbital characters of these bands 
along $\Gamma M$ and $\Gamma K$.  In both materials, along these in-plane 
directions, each band exhibits one dominant orbital character, and all 
of them disperse quadratically.

Interestingly, the higher-lying band along the in-plane directions has 
$p_z$ character in hcp Cd but $s$ character in hcp Ag.  This band 
inversion is reminiscent of the one noted in HgTe-CdTe quantum 
wells \cite{Bernevig2006}.  When the well is thin, the $s$-derived band 
lies higher than the $p$-derived band,  the orbital mixing is low, and 
the bands are quadratic with a finite gap at $\Gamma$.  With increasing 
thickness  of HgTe in the quantum well, the orbital characters start to mix, 
the gap decreases, and the bands start to become linearly dispersive.  
The gap vanishes and the bands cross each other linearly at a critical 
thickness, beyond which the higher-lying band exhibits more $p$ character. 
In this analogy, hcp Ag is like thin HgTe below the critical thickness and 
its $s$-type band lies above its $p_z$-type band.  Meanwhile, hcp Cd is 
analogous to thick HgTe above the critical thickness, because its
$p_z$-type band lies just above its $s$-type band. 

\begin{figure}
  \includegraphics[width=\columnwidth]{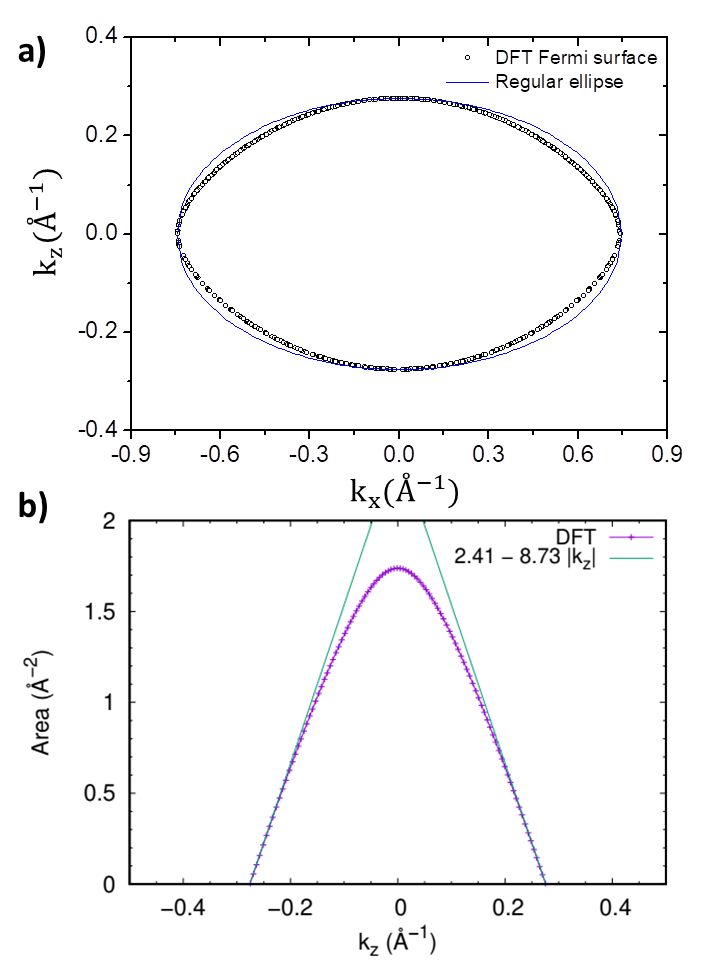}
  \caption{ a) The electron-like Fermi surface generated by the upper semi-Dirac band projected in the $k_x, k_z$ plane (empty circles). The blue solid line represents a regular ellipsoid whose axes are equally long. Note the difference. b) The cross section area  as a function of $k_z$.  Over a long distance, near the poles (that is $k_{z,min}$ and $k_{z,max}$), the area is linear in $k_z$, consistent with the linear dispersion of the semi-Dirac band from which this Fermi sheet derives.}
  \label{fig:area}
\end{figure}

Since the band structures of hcp Cd and hcp Ag have been calculated
using the same lattice parameters, the difference in the relative
position of their $s$- and $p_z$-type bands near the Fermi level is
due to the presence of an additional electron in Cd.  The extra
electron does not just rigidly fill up the unoccupied states in the
band structure when going from hcp Ag to hcp Cd, it also pulls down
the $s$-type states relative to the position of the $p_z$-type states.
We discuss the inversion of the relative positions of the $s$- and 
$p_z$-type states in hcp Ag due to tensile strain in the appendix.

Direct observation of the linear dispersion of these semi-Dirac bands 
in hcp Cd emerges as a task for future experiments employing angle-resolved 
photoemeission spectroscopy.  However, it has already a striking
experimental signature in the Sondheimer oscillations \cite{Sondheimer1950}  
of cadmium detected by experiments \cite{Grenier1966,Guo2024}.

In a metallic crystal  hosting ballistic electrons, galvonometric coefficients display oscillations, which are periodic in magnetic field. The period of these oscillations is given by \cite{Harrison,van2021sondheimer}:
\begin{equation}
  \Delta B = \frac{\hbar}{ed}\left(\frac{\partial A}{\partial k_z}\right),
  \label{Eq-Harrison}
\end{equation} 

Here,  $d$ is the thickness of the crystal in the direction perpendicular to the magnetic field, $e$ is the electron charge, $\hbar$ is the reduced Planck constant, $k_z$ is the component of the wave vector along the magnetic field, and $A$ is the cross section area of the Fermi sheet intersecting with a plane perpendicular to $k_z$.

In a perfect ellipsoid,  $\frac{\partial A}{\partial k_z}$ is not constant and smoothly varies as function of k$_z$. At the limit point (the apex) of the ellipsoid, it has a slightly higher degeneracy than everywhere else. With increasing number of states sharing an identical $\frac{\partial A}{\partial k_z}$, oscillations sharing the same periodicity become more robust. In our case, the electron-like pocket formed by the upper semi-Dirac band (Fig.~\ref{fig:bandst}d) is not an ellipsoid. Fig.~\ref{fig:area}a shows how the projection of this Fermi surface in the ($k_x$, $k_z$) plane is not a regular ellipsoid. Fig.~\ref{fig:area}b shows its cross section area as a function of $k_z$.  Along this orientation, the apices of this lens-like Fermi surface occur at $\pm0.28$ \AA$^{-1}$. The area is proportional to $k_z$ along a large fraction of the out-of-plane path in the Brillouin zone near these points, as expected by the linear dispersion in this direction.  The figure also shows a linear fit, which demonstrates that the quadratic component is negligible up to a distance of $0.09$ \AA$^{-1}$ towards the Brillouin zone center from the apices. The slope of the linear fit is $\frac{\partial A}{\partial k_z} = 8.73$ \AA$^{-1}$. 

This calculated slope is in excellent agreement with was found experimentally first by Grenier \etal \cite{Grenier1966} in a sample with $d=1.02$ mm. According to their data, $\delta B=0.0564\pm 0.0003$ T. Inserting these in Eq. \ref{Eq-Harrison}, yields $\frac{\partial A}{\partial k_z}=8.76$ \AA$^{-1}$. A recent experiment \cite{Guo2024} quantifying the periodicity across a forty-fold variation of thickness found an identical value of 8.76 \AA$^{-1}$.  Now, if the deviation from the ellipsoidal shape (see Fig.~\ref{fig:area}a) is neglected,   the expected period is the apex $\frac{\partial A}{\partial k_z}\big|_{F_z} = 2 \pi \frac{k_{F_r}^2}{k_{F_z}}$, which yields 12.58 \AA$^{-1}$. Such a large 44\% difference confirms the importance of the linear dispersion along $k_z$ in setting the period of the Sondheimer oscillations observed in this material. The agreement between theory and experiment implies that the slope of the out-of-plane linear dispersion near the Fermi level conforms to what is computed here. 

\section{Conclusions}

In summary, revisiting the electronic structure of hcp Cd using
first principles calculations demonstrates that it a hosts
deep-lying anti-crossing of semi-Dirac bands with linear out-of-plane dispersion and quadratic
in-plane dispersion.  We argue that this is driven by direction-dependent mixing of $s$ and $p_z$ states, and that hcp Cd can be likened to HgTe-CdTe near its critical thickness, except with the restriction of band hybridization along a single direction. The nonellipsoid electron pocket produced by the upper semi-Dirac band has a cross-section area  proportional to the out-of-plane wave vector along a significant fraction of the Brillouin zone path and is in excellent quantitative agreement with the period of Sondheimer oscillations.

\section{acknowledgements}
This work was supported by Grand Equipement National de Calcul Intensif
and  Tr\`es Grand Centre de Calcul (GENCI-TGCC) under grant no.\ A0150913028.

\appendix*

\begin{figure}[!htb]
  \includegraphics[width=\columnwidth]{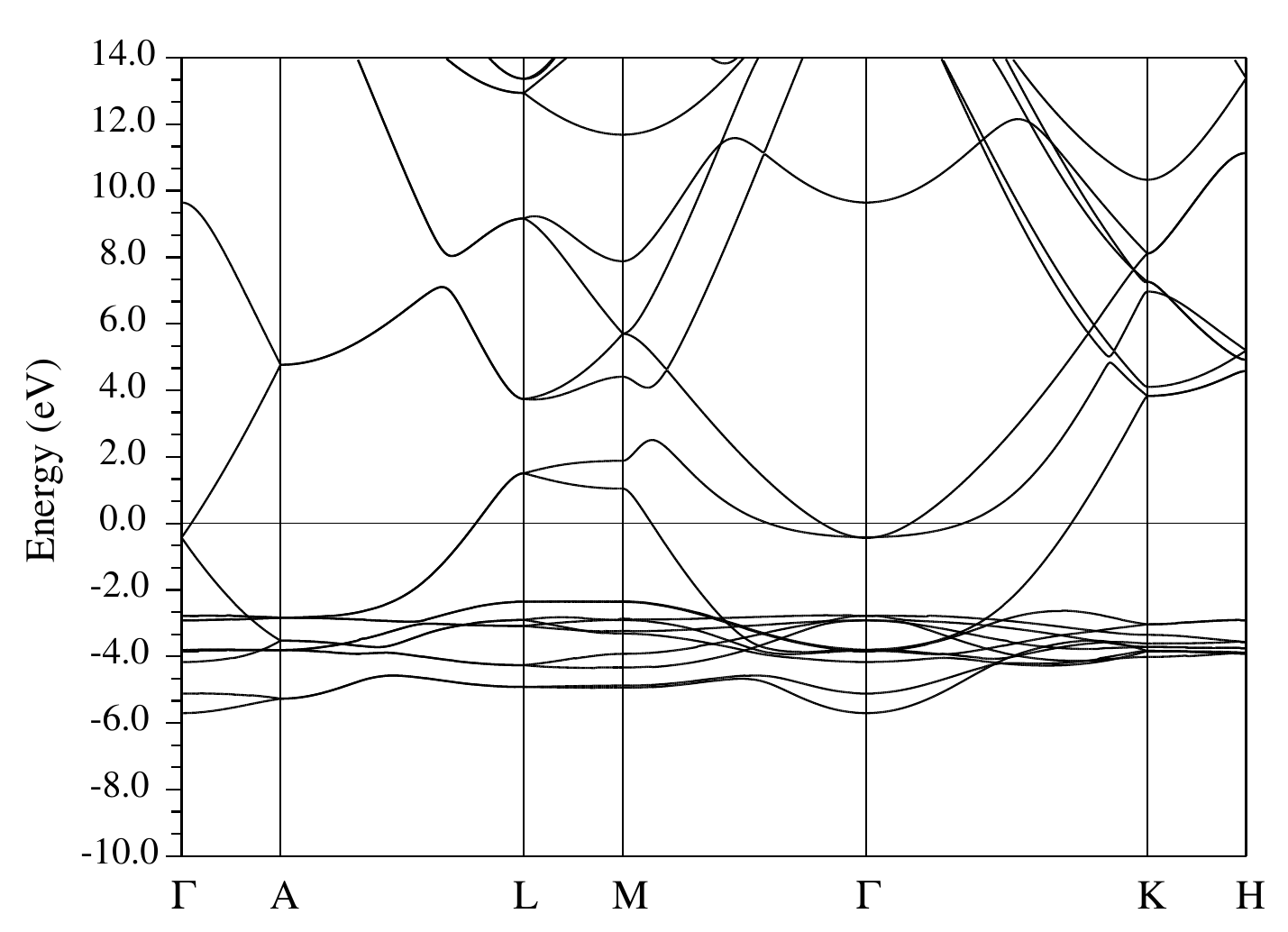}
 \caption{Calculated LDA band structure of hypothetical hcp Ag with the lattice parameter $c$ strained by 13\% compared to that of hcp Cd.  Semi-Dirac bands appear in hcp Ag when strained tensilely.}
\label{fig:bandstAgelong}
\end{figure}

\section{Engineering band dispersion with strain }

The relative position of the $s$- and $p_z$-type bands
near the Fermi level in hcp Ag and the hybridization between them can
be controlled by varying the lattice parameter $c$.  As $c$ is
increased, the gap at $\Gamma$ between the $s$ and $p_z$ states
decreases, and these bands become more linear near $\Gamma$ along the
out-of-plane direction.  The gap vanishes when $c$ is increased by
13\%, as can be seen in Fig.~\ref{fig:bandstAgelong} that shows the
band structure of hcp Ag at this value of the lattice parameter.  We
can also see that these bands converge at a semi-Dirac point, with
linear dispersion along the out-of-plane direction $\Gamma$-$A$ and
parabolic dispersion along the in-plane directions $\Gamma$-$M$ and
$\Gamma$-$K$ near the Brillouin zone center.

\begin{figure}[!htb]
\includegraphics[width=0.9\columnwidth]{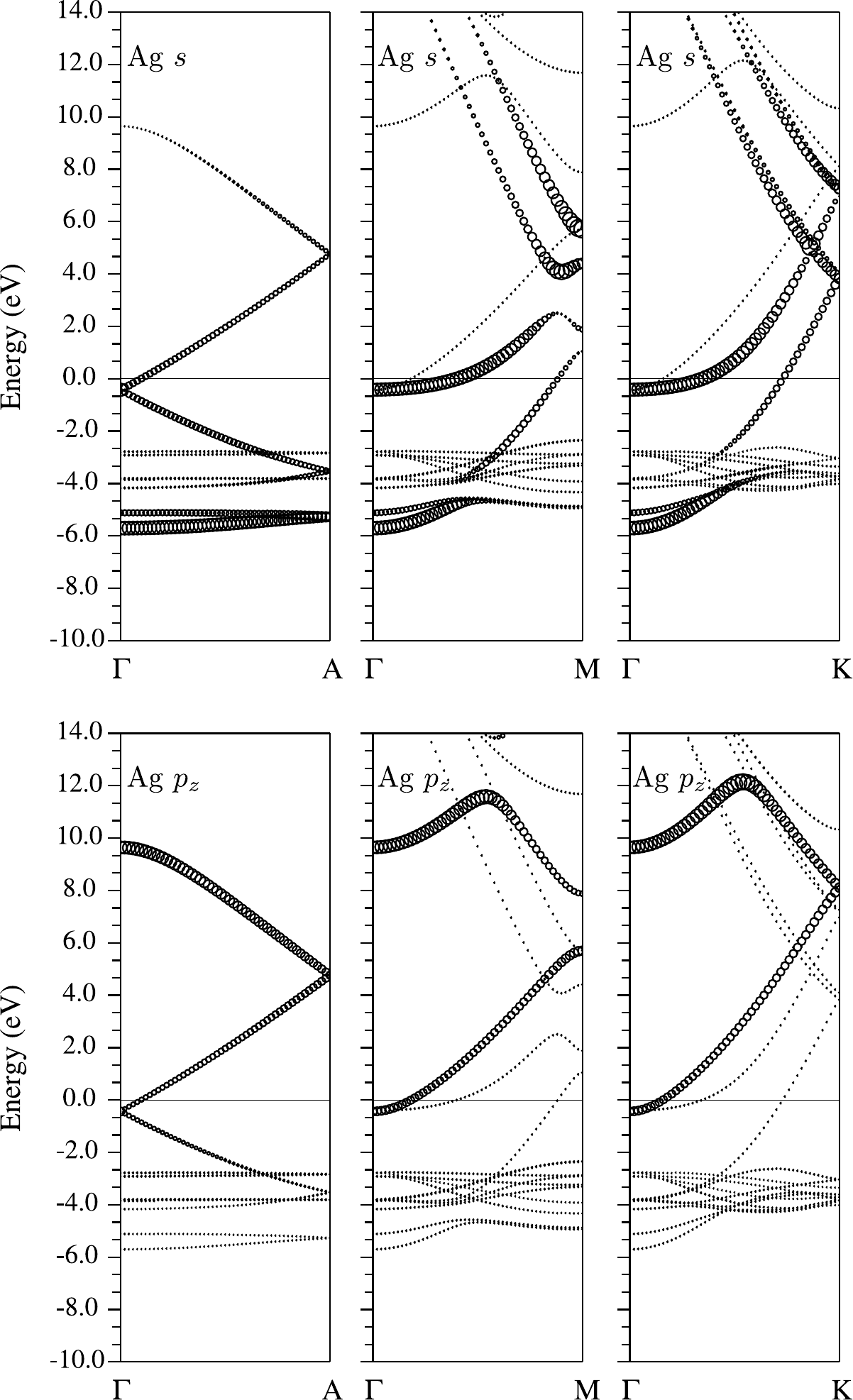}
  \caption{Same as in Fig.~\ref{fig:bandstAgelong}, but plotted as a function of $s$ and $p_z$ orbital characters along the out-of-plane $\Gamma A$ and inplane $\Gamma M$ and $\Gamma K$ directions.  The linearly dispersive bands in the out-of-plane  direction show mixed $s$ and $p_z$ characters, while the bands remain unmixed in the inplane directions.  The unmixed $s$-like  band lie below the unmixed $p_z$-like band, indicating that tensile strain inverts the relative positions of these bands. }
  \label{fig:bandchAgelong}
\end{figure}

Fig.~\ref{fig:bandchAgelong} shows the corresponding band structure
weighted with the $s$ and $p_z$ orbital contributions along the
out-of-plane and two inplane paths, and they are now similar to that
of hcp Cd.  Along the out-of-plane direction, the two semi-Dirac bands
show almost equal amount of $s$ and $p_z$ contribution.  There is very
little hybridization between these states in the inplane directions,
with the $p_z$-type band lying above the $s$-type band away from
$\Gamma$ as in the case of hcp Cd.  This confirms that
direction-dependent band hybridization driven by $s$-$p_z$ inversion
forms the semi-Dirac bands in these systems. In the case of HgTe-CdTe
quantum wells, the band inversion is caused by spin-orbit interaction.
Here, the inversion is tuned by interplanar distance. 
Of course, the strain value of 13\% is higly unphysical and
this hypothetical scenario mainly serves to provide an example of band 
inversion due to strain.

\bibliography{cadmium-semi-dirac}

\end{document}